\documentclass{article}
\usepackage{spconf}
\usepackage{amsmath,graphicx}
\usepackage{subfigure}
\usepackage{multirow}
\usepackage{bm}
\usepackage{threeparttable}
\usepackage{makecell}
\usepackage{booktabs}
\usepackage[table]{xcolor}

\title{Peak-First CTC: Reducing the Peak Latency of CTC Models \\by Applying Peak-First Regularization}
\name{Zhengkun Tian, Hongyu Xiang, Min Li, Feifei Lin, Ke Ding, Guanglu Wan}
\address{Meituan, Beijing, China}
\begin{document}
	\ninept
	\maketitle
	\begin{abstract}
		The CTC model has been widely applied to many application scenarios because of its simple structure, excellent performance, and fast inference speed. There are many peaks in the probability distribution predicted by the CTC models, and each peak represents a non-blank token. The recognition latency of CTC models can be reduced by encouraging the model to predict peaks earlier. Existing methods to reduce latency require modifying the transition relationship between tokens in the forward-backward algorithm, and the gradient calculation. Some of these methods even depend on the forced alignment results provided by other pretrained models. The above methods are complex to implement. To reduce the peak latency, we propose a simple and novel method named peak-first regularization, which utilizes a frame-wise knowledge distillation function to force the probability distribution of the CTC model to shift left along the time axis instead of directly modifying the calculation process of CTC loss and gradients. All the experiments are conducted on a Chinese Mandarin dataset AISHELL-1. We have verified the effectiveness of the proposed regularization on both streaming and non-streaming CTC models respectively. The results show that the proposed method can reduce the average peak latency by about 100 to 200 milliseconds with almost no degradation of recognition accuracy.
		
	\end{abstract}
	
	\section{Introduction}
	The Connectionist Temporal Classification (CTC) models have achieved great success in speech recognition\cite{graves2006connectionist,hannun2014deep,amodei2016deep,salazar2019self,li2019jasper,kriman2020quartznet}. Compared with the transducer models\cite{graves2012sequence,graves2013speech,he2018streaming} and the attention-base encoder-decoder models\cite{chan2016listen,vaswani2017attention}, the structure of the CTC model is very simple, including only two parts: an acoustic encoder and an output linear projection layer. Relying on the forward-backward algorithm, the CTC model can optimize all alignment paths and directly model the conversion relationship from acoustic sequences to the corresponding text sequences. The simple structure makes the CTC models have an unparalleled inference speed advantage over other end-to-end models.
	
	The CTC model introduces a blank token($\phi$) to model meaningless audio fragments (including mute, pause, noise and so on) or duplicate non-blank tokens. This property makes the probability distribution of CTC models generate many probability peaks, where each peak represents a non-blank token\cite{graves2006connectionist,senior2015acoustic,tian2020spike}. The earlier the probability peak appears, the lower latency in recognizing the corresponding token. Based on the previous observations, the position of the peaks is closely related to the model structure. The model with stronger context modeling ability, the earlier its peak is. For the streaming ASR model, due to the lack of future acoustic information, the peak position predicted by the model is usually a little later than the non-streaming models.
	
	How to make the CTC or transducer models able to reduce the recognition latency by predicting non-blank tokens earlier has become a hot issue in recent years\cite{senior2015acoustic,yu2021fastemit,li2021better,shinohara22_interspeech,tian2022bayes,inaguma2021stableemit,kim2017joint}. There are many ways to alleviate this problem. Firstly, constrained alignment training limits the path search space of forward-backward algorithm by forcing the distance between any alignment path and "ground truth" alignment to be no more than a threshold\cite{senior2015acoustic,shinohara22_interspeech,inaguma2021stableemit}. This method requires some external tools to generate forced alignment paths in advance. Secondly, the FastEmit method reduces the recognition latency by decomposing the path through each node in the forward-backward path space and encouraging the model to preferentially predict non-blank tokens\cite{yu2021fastemit,li2021better,tian21_interspeech}. Thirdly, the Bayes Risk CTC (BRCTC) \cite{tian2022bayes} divides all feasible paths into different groups and assigns different Bayesian risk values to the different path groups. The risk value is set to encourage the model to focus on optimizing low-latency paths. The above methods are complex to implement, and it is necessary to modify the forward-backward calculation process and the gradient calculation of CTC models. Finally, \cite{kim2021reducing} proposed a new scheme named self alignment which constructs a low-latency transition path based the self-alignment result, and takes its negative log-probability as a regularization term to optimize. This method does make no any modification on the calculation of CTC loss, but requires the model to continuously generate self-alignment paths online, which may consume a lot of computing resources.
	
	Different from the transducer models, the alignment paths of CTC model can only move along the time axis. The movement of peaks predicted by the CTC model is closely related to the movement of its output probability distribution. If we move the output probability distribution to the left along the time axis, the peak positions will move with it, and the latency of the corresponding path will also be reduced. Based on the above observations, we propose a new method name peak-first regularization(PFR) to reduce the recognition latency. The proposed method introduces a simple function to shift the output probability distribution, which does not modify the gradient calculation and does not rely on additional forced alignment methods. The regularization term is combined with a weight hyper-parameter and the original CTC loss. The latency can be reduced in varying degrees by setting different weights. During the training, the probability distribution of the CTC model will not continue to shift left, but seeks a balance between the peak latency and recognition accuracy. Our experiments are conducted on a Chinese Mandarin dataset AISHELL1. The results show that the proposed peak-first regularization method can effectively reduce the peak latency, and even achieves a small increase in the recognition accuracy.
	
	The remainder of this paper is organized as follows. Section 2 will describe the classical CTC method and our proposed Peak-First CTC model. Section 3 presents the experimental setup and results. The conclusions and future work will be given in Section 4. 
	
	\begin{figure*}[t]
		\centering
		\subfigure[The CTC Alignment Path Space]{
			\centering
			\label{fig:pfgrid}
			\includegraphics[width=0.39\linewidth]{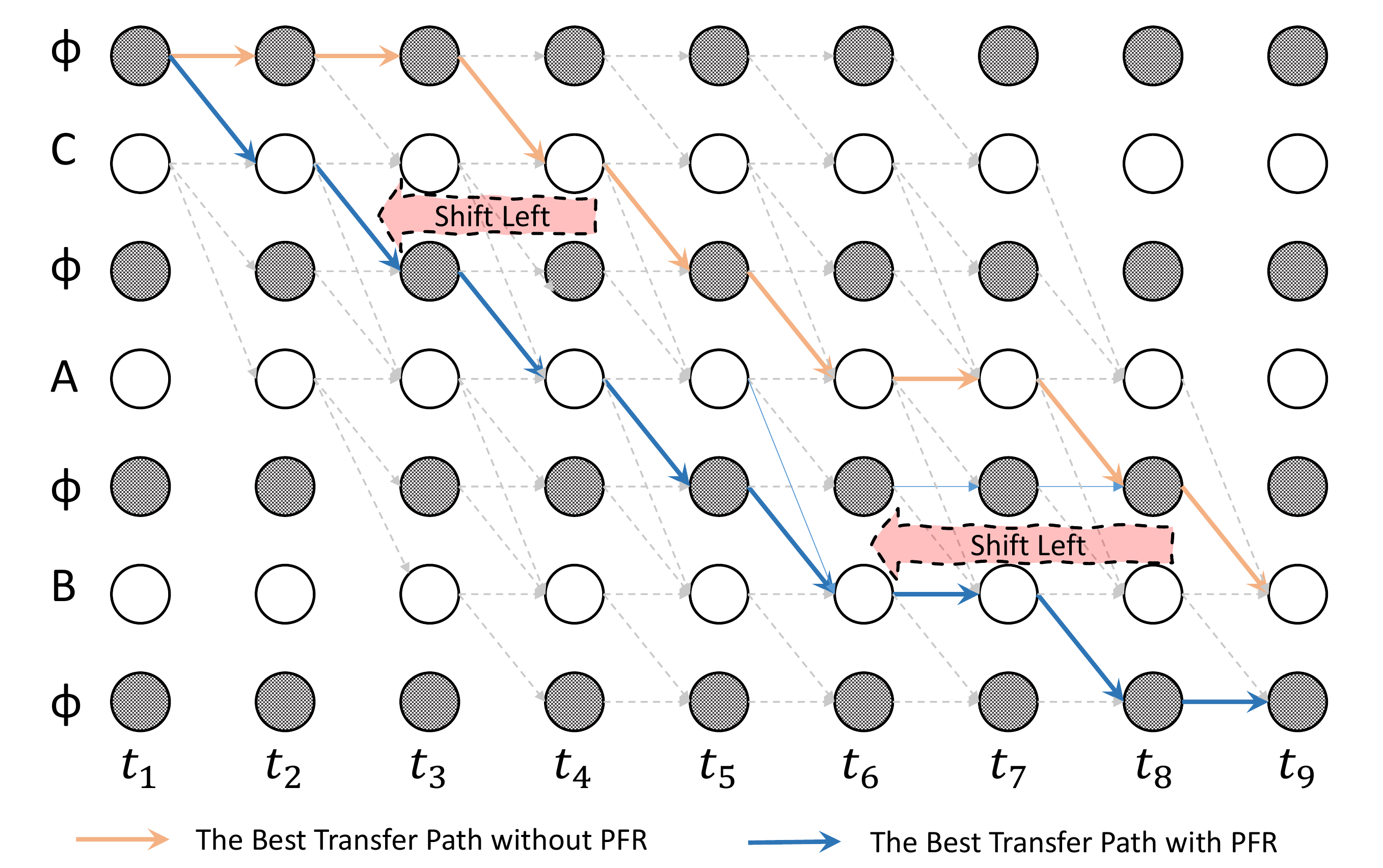}}
		\subfigure[The Calculation of Peak-First Regularization]{
			\centering
			\label{fig:pfloss}
			\includegraphics[width=0.58\linewidth]{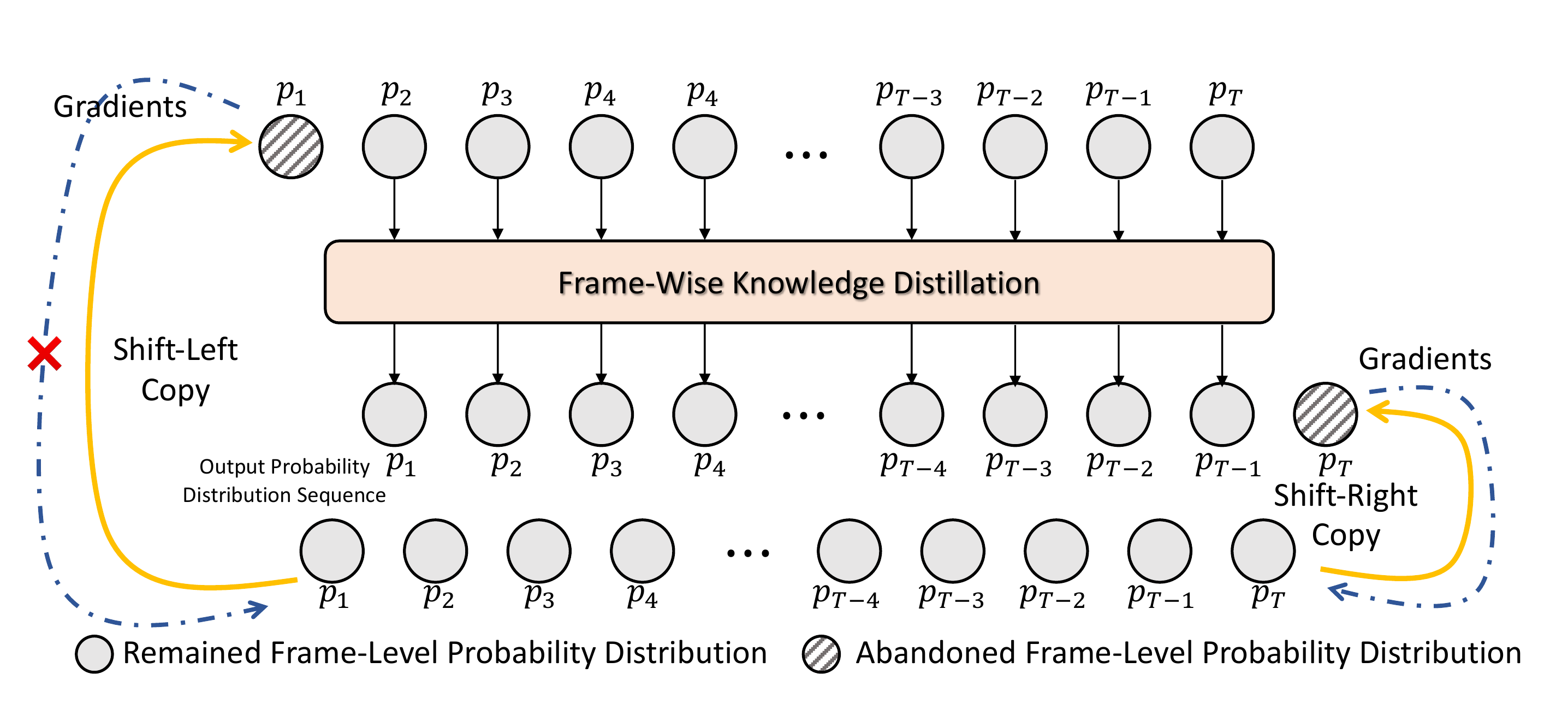}}
		\caption{(a) illustrates the alignment path space of CTC model. Each node represents the probability that the CTC model predicts the corresponding token at time $t$. All points connected by a line form an alignment path. The position of non-blank tokens on the blue line appears earlier than the tokens on the orange line, so the blue alignment path has lower recognition latency. (b) introduces the calculation of peak-first regularization. The regularization utilizes a frame-wise knowledge distillation method to encourage each frame output of the model to learn the probability distribution from its right adjacent frame.}
		 \vspace{-10pt}
	\end{figure*}
	
	\section{Peak-First CTC}
	\subsection{Connectionist Temporal Classification}
	The connectionist temporal classification method is proposed to model the conversion between the acoustic feature sequence $\textbf{x}$ and the corresponding text sequence $\textbf{y}$, which maximizes the posterior probability $P(\textbf{y}|\textbf{x})$. In order to obtain audio-text pairs with equal length, the CTC model introduces a blank token (represented by $\phi$) and allow continuous duplicate tokens to exist. For any acoustic sequence with a length of $T$, there are many feasible text sequences, each of which is a feasible alignment path (represented by $\pi$), as depicted by the gray dotted line in Fig.\ref{fig:pfgrid}. The CTC model aims to minimize the negative log-probability:
	\begin{equation}
		\mathcal{L}_{CTC}=-\log P(\textbf{y}|\textbf{x})=-\sum_{\pi \in \mathcal{B}^{-1}(\textbf{y})}P(\pi|\textbf{x})
	\end{equation}
	where $\mathcal{B}^{-1}(\textbf{y})$ is the alignment path space generated based on the ground-truth text sequence $\textbf{y}$. The forward-backward algorithm is adopted to sum the probabilities of all alignment paths in the space $\mathcal{B}^{-1}(\textbf{y})$. To facilitate calculation, a new token sequence $\textbf{y}'$ ($|\textbf{y}'|=U'$) is constructed, with blank token added to the beginning and the end and inserted between any two adjacent tokens in sequence $\textbf{y}$($|\textbf{y}|=U$). The length of the two sequences satisfies the following relationship: $U'=2U+1$.
	
	The \textit{Forward variable} $\alpha(t,u)$ is defined as the sum of the probabilities of all alignment paths from the two starting positions to the $u$-th $(0\leq u \leq U')$ position in the sequence $\textbf{y}'$ at time $t(1\leq t \leq T)$. The variables can be calculated recursively:
	\begin{equation*}
		\alpha(t,u)=\begin{cases}
			p_{y'_u}^t\sum_{i=u-1}^{u}\alpha(t-1,i),\text{if}\quad y'_u=\phi \quad\text{or}\quad y'_{u}=y'_{u-2}\\
			p_{y'_u}^t\sum_{i=u-2}^{u}\alpha(t-1,i),\text{otherwise}
		\end{cases}
	\end{equation*}
	where $p_{y'_u}^t$ is the probability of token $y'_u$ predicted by the model at time $t$.
	
	The \textit{Backward variable} $\beta(t,u)$  is defined as the sum of the probabilities of all alignment paths from the the $u$-th $(0\leq u \leq U')$ position in the sequence $\textbf{y}'$ at time $t(1\leq t \leq T)$ to the two end positions.
	\begin{equation*}
		\beta(t,u)=\begin{cases}
			\sum_{i=u}^{u+1}\beta(t+1,i)p_{y'_i}^{t+1},\text{if}\quad y'_u=\phi \quad\text{or}\quad y'_{u}=y'_{u+2}\\
			\sum_{i=u}^{u+2}\beta(t+1,i)p_{y'_i}^{t+1},\text{otherwise}
		\end{cases}
	\end{equation*}
	
	The loss function can be re-expressed by forward and backward variables:
	\begin{equation}
		\mathcal{L}_{CTC}=-\log P(\textbf{y}|\textbf{x})=-\log \sum_{u=1}^{U'}\alpha(t,u)\beta(t,u)
	\end{equation}
	The gradient of any output probabilities also be expressed by the forward and backward variables.
	
	\subsection{Peak-First Regularization}
	The alignment paths of CTC models unfold along the time axis. As show in the Fig.\ref{fig:pfgrid}, all non-blank tokens points on the blue alignment path appear earlier than points on the orange one. This also means that the blue path has a lower recognition latency. It's easy to get an intuitive observation that low-latency paths are further to the left on the time axis. Therefore, we conjecture that the peak latency can be reduced by making the probability distribution of CTC model shift left.
	
	We propose a peak-first regularization (PFR) method, which encourages the CTC model to learn the probability distribution from its right adjacent frame. It does not directly modify the path transfer relationship in the forward-backward path search space, nor does it rely on the forced alignment or other methods to provide low-latency ground-truth paths. The calculation of PFR depends on a frame-wise knowledge distillation function:
	\begin{equation}
		\mathcal{L}_{PFR}=\sum_{t=1}^{T-1}\bm{p}^{t+1}\log\frac{\bm{p}^{t+1}}{\bm{p}^{t}}=\sum_{t=1}^{T-1}\sum_{k=1}^{V} p_{k}^{t+1}\log\frac{p_{k}^{t+1}}{p_{k}^{t}}
	\end{equation}
	where $\bm{p}^{t}$ is the output probability distribution over the vocabulary, and $V$ is the size of vocabulary. In the distillation process, the too high token probability predicted by CTC models may inhibit the learning of other token. Therefore, we introduce a temperature coefficient $\tau$ to smooth the probability distribution.
	\begin{equation}
		p_{k}=\frac{exp(o_k^t/\tau)}{\sum_{1 \leq k' \leq V}exp(o'^t_{k'}/\tau)}
	\end{equation}
	In the following experiments, the default $\tau$ is 10.0.
	
	\subsection{Model Optimization and Gradient Analysis}
	The joint training process can be expressed as follows:
	\begin{equation}
		\mathcal{L}=\mathcal{L}_{CTC} + \lambda \mathcal{L}_{PFR}
	\end{equation}
	where $\lambda$ is the weight value of peak-first regularization. During the training, the output probability distribution will not continuously shift to the left, leading to the collapse of model training. Under the appropriate weight setting, the model will reach a balance between the recognition accuracy and latency. At this time, the distribution will no longer move and the value of  PFR regularization term will also reach a stable state.
	
	We can also analyze the reason why the recognition latency can be reduced by moving the probability distribution from the perspective of gradient calculation. The gradient of CTC loss against the output probability $p^{t}_{k}$ can be written as:
	\begin{equation}
		\frac{\partial \mathcal{L}_{CTC}}{\partial p^{t}_{k}}=-\frac{G(t,k)}{p^{t}_{k}}
	\end{equation}
	where $G(t, k)=\frac{\sum_{u\in \{u: y'_u=k\}}\alpha(t,u)\beta(t,u)}{P(\textbf{y}|\textbf{x})}$. After the regularization term is adopted, the gradient will be updated as follows:
	\begin{equation}
		\frac{\partial \mathcal{L}}{\partial p^{t}_{k}}=-\frac{G(t,k)+\lambda p^{t+1}_{k}}{p^{t}_{k}}
	\end{equation}
	The gradient of the current position is closely related to the probability value $p^{t+1}_{k}$ of the next frame. The probability value $p^{t+1}_{k}$ is larger when the CTC model predicts a peak of $k$-th token at time $t+1$. This will increase the gradient of the current frame and encourage the current frame to change rapidly, thereby making the peak move forward. On the contrary, if the probability of the next frame is very small, there is almost no effect on the gradient of the current frame.
	
	\begin{table*}[t]
		\centering
		\caption{Comparison of different models and different PFR weights $\lambda$. APL denotes the average peak latency.}
		\begin{tabular}{cccccccccc}
			\toprule
			\multirow{2}{*}{Model} & \multirow{2}{*}{\makecell[c]{Weight $\lambda$}}  & \multicolumn{4}{c}{Dev}&\multicolumn{4}{c}{Test}\\
			\cline{3-10}
			
			& & CER & APL(ms) & PR50(ms) & PR90(ms) & CER & APL(ms) & PR50(ms) & PR90(ms) \\
			\hline
			
			\multirow{6}{*}{\makecell[c]{Non-Streaming\\Models}} & Baseline & 
			5.38 & -187.24 & -300 & -230 & 5.85 &	-202.44 & -300 &	-220 \\
			
			\multirow{6}{*}{} & $0.1$ & 5.28 & -214.82	&-330	&-250 &5.71 &-230.01	&-330 &-240 \\
			
			\multirow{6}{*}{} & $0.3$ & 5.36 &-275.45 & -390	& -310 & 5.79 & -289.88	& -390	& -300 \\
			
			\multirow{6}{*}{} & $0.5$ & 5.41 & -337.35	& -460	& -370 & 5.83 & -351.48	& -450	& -350 \\
			
			\multirow{6}{*}{} & $0.7$ & 5.43 & -355.61	& -470 & -380 & 5.97 & -369.58	&-470 & -360\\
			
			\multirow{6}{*}{} & $0.9$ &5.59	& -403.33	& -510 & -420	& 6.00	& -419.29	& -510	& -400\\
			\hline
			\multirow{6}{*}{\makecell[c]{Streaming\\Models\\(510ms)}} & Baseline & 
			6.43 & 400.23 &	280 & 360 & 7.03 & 384.95 &	280 & 360 \\
			
			
			\multirow{6}{*}{} & $0.5$ &6.25	&361.16	&240	& 320	& 6.86	&346.09	& 240	&330\\
			
			
			
			\multirow{6}{*}{} & $1.0$ &6.16	&343.39	&220	&300 &6.81	&328.44	&220	&310\\
			
			\multirow{6}{*}{} & $2.0$ &6.17	&326.23	&210	&280 &6.85	&311.62	&210	&290\\
			
			\multirow{6}{*}{} & $3.0$ &6.21	&297.10	&180	&260 &6.84	&283.22	&180	&270\\
			
			\multirow{6}{*}{} & $5.0$ &6.52	&217.81	&90	&190 &7.22	&206.44	&90	&200\\
			
			\toprule
		\end{tabular}
		\label{tab:res}
	\end{table*}
	
	\section{Experiments and Results}
	
	\subsection{Dataset}
	In this work, all experiments are conducted on a public Chinese Mandarin speech corpus AISHELL-1\footnote{https://openslr.org/33/}. The training set contains
	about 150 hours of speech (120,098 utterances) recorded by 340 speakers. The development set contains about 20 hours (14,326 utterances) recorded by 40 speakers. And about 10 hours (7,176 utterances / 36109 seconds) of speech is used as the test set. The speakers of different sets do not overlap.
	
	\subsection{Experimental Setup}
	For all experiments, we use 83-dimensional FBANK with pitch features computed on a 25ms window with a 10ms shift. We choose 4234
	characters (including a blank token \texttt{<$\phi$>} and an unknown token \texttt{<UNK>}) as modeling units.
	
	Based on the transformer architecture\cite{vaswani2017attention}, we have built  a streaming and non streaming model respectively. Both of them contain 12 encoder blocks. The models apply a 2D convolution front end, which utilizes two-layer time-axis CNN with ReLU activation, stride size 2, channels 256, and kernel size 3 \cite{watanabe2018espnet}. Each encoder block consists of a multi-head self-attention layer and a feed-forward layer. There are 4 heads in multi-head self-attention layer. Both the output size of the multi-head self-attention and the feed-forward layers are 256. The feed-forward layers adopt GELU\cite{hendrycks2016gaussian} as the activation function and have 2048 hidden units. We utilize an Adam optimizer with warm-up steps 25000 and the learning rate scheduler reported in \cite{vaswani2017attention}. After 80 epochs, we average the parameters saved in the last 20 epochs. We also use 3-way speed perturbation and the time-masking and frequency-masking \cite{park2019specaugment} for data augmentation. For the streaming models, we utilize the frame-level mask to shield the long-range future information that the self-attention layers focus on. The streaming models can only pay attention to the following acoustic information of 510 milliseconds.
	
	We use the character error rate (CER) to evaluate the performance of different models. Average token latency and partial recognition latency\cite{yu2021fastemit} are used to evaluate the recognition latency. Average peak latency is the average time difference between the last frame of speech corresponding to each peak and the peak emitted time. The partial recognition latency represents the difference between the time when the speaker stops speaking and the time when the last token is emitted. The ground-truth alignment result is provided by a GMM model\footnote{The GMM model is trained according to the following script: https://github.com/kaldi-asr/kaldi/tree/master/egs/aishell2/s5/run.sh}.
	\subsection{Results}
	
	\begin{figure*}[t]
		\centering
		\subfigure[Comparison of Different Non-Streaming Models]{
			\centering
			\label{fig:ns}
			\includegraphics[width=0.48\linewidth]{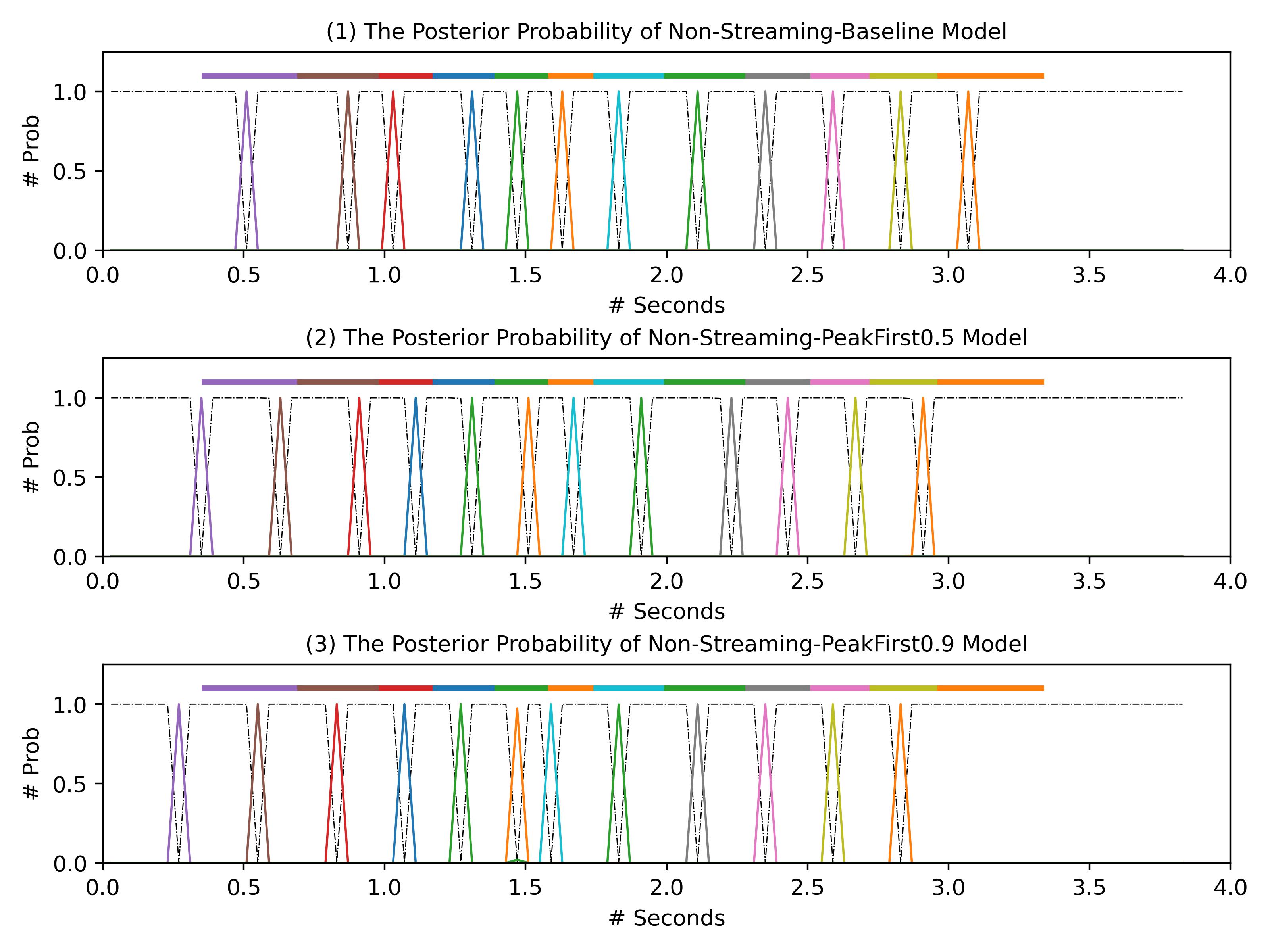}}
		\subfigure[Comparison of Different Streaming Models]{
			\centering
			\label{fig:s}
			\includegraphics[width=0.48\linewidth]{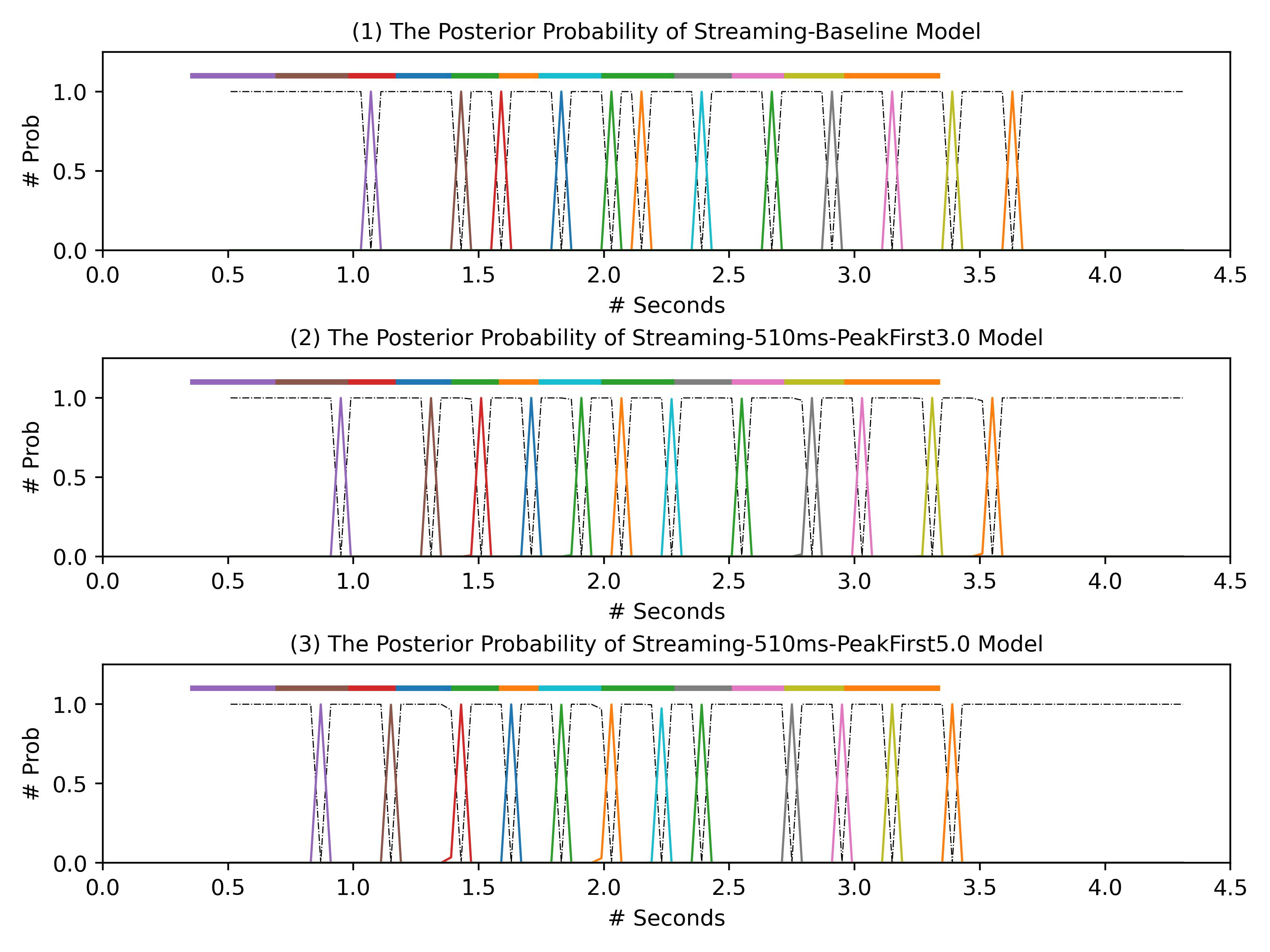}}
		\vspace{-10pt}
		\caption{Visualization Analysis of Different Models and Peak-First Regularization Weights. The ID of above utterance is \texttt{BAC009S0764W0135}, which comes from the AISHELL test set. The top thick solid line represents the character-level forced alignment result provided by the GMM model. The peaks represent output probability distribution of the non-blank token predicted by CTC models.}
		\label{fig:peak}
		\vspace{-15pt}
	\end{figure*}
	
	\subsubsection{Compare the effects of peak-first regularization term on recognition latency}
	We first analysis the effects of our proposed PFR regularization  on the recognition latency from three views. We report the average peak latency, 50-th (PR50) and 90-th (PR90) percentile recognition latency of all utterances in the test set. For the baseline model, the weight $\lambda$ is 0.0. All results are recorded in the Tab.\ref{tab:res}. 
	
	It is easy to find that the average peak latency and partial recognition latency of non-streaming models are negative, while the latency of streaming models are positive. The reason for this phenomenon is that the non-streaming models does not need to consider the latency caused by the following acoustic information, and the predicted peaks tend to be located in the pronunciation range of each non-blank token. By contrast, the latency (510ms) caused by the limited future acoustic context is also taken into account. This makes the peaks of the streaming model move backward, which exceeds their corresponding acoustic range. This analysis also can be verified from the Fig.\ref{fig:peak}.
	
	The weight $\lambda$ of the regularization term can be utilized to adjust the recognition latency. As shown in Tab.\ref{tab:res}, the non-streaming model with weight $0.9$ can achieve the best latency performance on the test set, which reduces the average peak latency by 216.85ms, PR50 by 210ms and PR90 by 180ms compared the baseline model. For the streaming models, when the weight is $5.0$, the model can achieve an average peak latency of 217.81ms on the test set, which is 178.51 ms less than the baseline. And the latency of PR50 and PR90 can reduce by 190ms and 160ms respectively. The greater the weight of regularization term, the lower  latency that the model can obtain. As depicted in the Fig.\ref{fig:peak},  the peaks with greater weights are further left on the time axis. However, a too large weight setting may break the balance between the recognition accuracy and latency, resulting in a degradation in model recognition performance. Considering the difference between the acoustic information range obtained by the non-streaming and streaming models, their optimal weight settings are also different. Overall, under the appropriate weight setting, the peak-first regularization can reduce the recognition latency by about 100 to 200 milliseconds.
	\vspace{-10pt}
	
	\subsubsection{Compare the effects of peak-first regularization term on recognition accuracy}
	\vspace{-5pt}
	As show in the Tab.\ref{tab:res}, it's obvious that the peak-first regularization has a little positive effect on improving the recognition accuracy. Under most weight parameter settings, both of the non-streaming and streaming model can achieve some improvement on the recognition accuracy. We guess there are two reasons for this phenomenon. One one hand, the peak-first regularization has no any modification on the forward-backward algorithm and gradients calculation, which does not destroy the modeling ability of the model itself. On the other hand, the peak-first regularization encourages each frame of the output sequence to learn the probability distribution from its right adjacent frame. The right output frame has more future acoustic information the left one, which may play a role in distilling knowledge.

	\section{Conclusions and Future works}
	We noticed that the alignment path space of the CTC model exists in the output probability distribution. As the output probability distribution moves to the left, the predicted peak position will also move along the time axis, which will reduce the recognition latency of the CTC model. Based on this observation, we propose a new method named peak-first regularization to encourage the CTC model predict the non-blank peaks earlier. The proposed method does not make any modification on the forward-backward algorithm and gradients calculation, and not rely on any forced alignment paths generated by external models. We conduct all the experimental on an open Chinese Mandarin dataset AISHELL-1. And the results show that both the non-streaming and streaming CTC model with peak-first regularization can reduce the recognition latency by about 100 to 200 milliseconds and have almost no degradation in recognition accuracy. In the future work, we will verify the effectiveness of our method on other language datasets and try to further reduce the recognition latency of the CTC models.

\vfill\pagebreak
	
\bibliographystyle{IEEEbib}
\bibliography{refs}
	
\end{document}